\documentclass[12pt,a4paper]{article}
\usepackage{jhep-mod}
\usepackage{bm}
\usepackage{amssymb, amsmath, amsthm, physics}
\usepackage{mathrsfs}
\usepackage{multicol}
\usepackage{float, array, tabu}
\usepackage{tensor}
\usepackage{tocloft}
\usepackage[utf8]{inputenc}
\usepackage{enumerate}
\usepackage{bigints}
\usepackage{xcolor}
\usepackage{appendix}
\usepackage{graphicx}
\usepackage{tikz}
\usepackage{setspace}
\usepackage{cancel}
\allowdisplaybreaks
\usepackage[normalem]{ulem}
\definecolor{purple}{rgb}{1,0,1}
\definecolor{lime}{HTML}{A6CE39} 

\DeclareMathOperator\diag{diag}
\DeclareMathOperator\sign{sign}
\def\d{{\mathrm{d}}}
\def\O{{\mathcal{O}}}
\def\A{{\mathcal{A}}}
\definecolor{darkgreen}{rgb}{0.0, 0.2, 0.13}
\definecolor{darkjunglegreen}{rgb}{0.1, 0.14, 0.13}

\definecolor{lime}{HTML}{A6CE39}
\newcommand{\orcidicon}{%
	\begin{tikzpicture}
	\draw[lime, fill=lime] (0,0) 
		circle [radius=0.16] 
		node[white] {{\fontfamily{qag}\selectfont \tiny ID}};
	\draw[white, fill=white] (-0.0625,0.095) 
		circle [radius=0.007];
	\end{tikzpicture}%
}
\newcommand{\orcid}[1]{{\href{https://orcid.org/#1}{\orcidicon}}}
\newcommand\orcidAlex{{\href{https://orcid.org/0000-0002-1763-3563}{\orcidicon}}}
\newcommand\orcidMatt{{\href{https://orcid.org/0000-0003-1088-6485}{\orcidicon}}}

\title{Charged black-bounce spacetimes}

\author{Edgardo Franzin\orcid{0000-0002-5705-5550}$^{a,b,c}$,}
\author{Stefano Liberati\orcid{0000-0002-7632-7443}$^{a,b,c}$,}
\author{Jacopo Mazza\orcid{0000-0002-0079-2234}$^{a,b,c}$, \break}
\author{Alex Simpson\orcidAlex$^d$,}
\author{{\sf{and}} Matt Visser\orcidMatt$^d$}

\affiliation{$^a$ SISSA, International School for Advanced Studies,\\via Bonomea 265, 34136 Trieste, Italy}
\affiliation{$^b$ INFN, Sezione di Trieste,\\via Valerio 2, 34127 Trieste, Italy}
\affiliation{$^c$ IFPU, Institute for Fundamental Physics of the Universe,\\via Beirut 2, 34014 Trieste, Italy}
\affiliation{$^d$ School of Mathematics and Statistics, Victoria University of Wellington,\\PO Box 600, Wellington 6140, New Zealand.}

\emailAdd{efranzin@sissa.it}
\emailAdd{liberati@sissa.it}
\emailAdd{jmazza@sissa.it}
\emailAdd{alex.simpson@sms.vuw.ac.nz}
\emailAdd{matt.visser@sms.vuw.ac.nz}

\abstract{Given the recent development of rotating black-bounce--Kerr spacetimes,
for both theoretical and observational purposes it becomes interesting to see whether it might be possible to construct black-bounce variants of the entire Kerr--Newman family.  
Specifically, herein we shall consider black-bounce--Reissner--Nordstr\"om and black-bounce--Kerr--Newman spacetimes as particularly simple and clean everywhere-regular black hole ``mimickers'' that deviate from the Kerr--Newman family in a precisely controlled and minimal manner, and smoothly interpolate between regular black holes and traversable wormholes. 
While observationally the electric charges on astrophysical black holes are likely to be extremely low, $|Q|/m \ll 1$, introducing any non-zero electric charge has a significant theoretical impact. In particular, we verify the existence of a Killing tensor (and associated Carter-like constant) but without the full Killing tower of principal tensor and Killing--Yano tensor, also we discuss how, assuming general relativity, the black-bounce--Kerr--Newman solution requires an interesting, non-trivial matter/energy content.

}


\arxivnumber{2104.11376}

\begin{document}

\maketitle

\section{Introduction}\label{S:intro}

Given the demonstrated existence of the black-bounce--Schwarzschild~\cite{Simpson:2018, Simpson:2019, Lobo:2020bb, Lobo:2020, Bronnikov:2005, Bronnikov:2006}, and the black-bounce--Kerr~\cite{Liberati:2021, Shaikh:2021, lima_junior_mistaken_2021} geometries, it is intuitive to suspect that analogous black-bounce variants of both the Reissner--Nordstr\"om (RN) and Kerr--Newman (KN) spacetimes will exist~\cite{Yang:2021};  and that they would be amenable to reasonably tractable general relativistic analyses.

Specifically, in ref.~\cite{Liberati:2021}, three of the current authors used the Newman--Janis procedure~\cite{newman_note_1965,erbin_janis-newman_2017, Rajan:2016} to transmute the original black-bounce--Schwarzschild spacetime~\cite{Simpson:2018,Simpson:2019,Lobo:2020bb,Lobo:2020} into an axisymmetric rotating version. Herein, we shall propose that the procedure by which one obtains a ``black-bounce'' variant from a known pre-existing solution in either axisymmetry or spherical symmetry can be simplified, and in doing so we advocate for two new candidate spacetimes: a spherically symmetric black-bounce with electrical charge (black-bounce--Reissner--Nordstr\"om), as well as the axisymmetric rotating equivalent (black-bounce--Kerr--Newman).

Given any spherically symmetric or axisymmetric geometry equipped with some metric $g_{\mu\nu}$ which possesses a curvature singularity at $r=0$, the proposed procedure is explicitly designed to transmute said geometry into a globally regular candidate spacetime whilst retaining the manifest symmetries. In standard $t,r,\theta,\phi$ curvature coordinates, the procedure is simply as follows:
\vspace{-5pt}
\begin{itemize}
\itemsep-3pt
\item Leave the object $\d r$ in the line element undisturbed.
\item Whenever the metric components $g_{\mu\nu}$ have an explicit $r$-dependence, replace the $r$-coordinate by $\sqrt{r^2+\ell^2}$, where $\ell$ is some length scale (typically associated with the Planck length).
\end{itemize}
\vspace{-5pt}
Leaving the object $\d r$ unchanged implies that the $r$-coordinate still performs an identical role to the curvature $r$-coordinate in terms of the spatial slicings of the spacetime, and ensures that we are not simply making a coordinate transformation. As we shall shortly see, replacing $r\rightarrow\sqrt{r^2+\ell^2}$ has the advantage of `smoothing' the geometry into something which is globally regular.

Apart from regularity, these spacetimes exhibit the interesting feature that they are a 1-parameter class of geometries smoothly interpolating between standard general relativity electrovac black holes and traversable wormholes~\cite{Morris-Thorne,MTY,Visser-book,Visser:1989a,Visser:1989b}.
In particular, the classical energy conditions will have non-trivial behaviour in these spacetimes~\cite{Visser-book, Visser:1989a,Visser:1989b, Visser:2003,Kar:2004,EC-survey-LNP, gvp4, gvp2, gvp1, Martin-Moruno:2013, twilight, Hochberg:1998,Visser:protection}. 
Furthermore, since these geometries can be fine-tuned to be arbitrarily close to the usual Kerr family, they can serve as black hole ``mimickers'' potentially of interest to the observational community~\cite{LISA,Carballo-Rubio:2018,Damour:2007,Carballo-Rubio:viability,Carballo-Rubio:complete,Pandora,small-dark-heavy,Visser:bh-in-GR,Mazur:2001,Mazur:2004,stable-gravastars,Visser:2014}.

\section{Black-bounce--Reissner--Nordstr\"om geometry}

To begin with, let us consider the Reissner--Nordstr\"om solution to the electrovac Einstein equations of general relativity expressed in terms of standard $t,r,\theta,\phi$ curvature coordinates and geometrodynamic units (we shall use everywhere a $(-,+,+,+)$ signature) 
\begin{equation}
    \d s^2 = -f_\text{RN}(r)\d t^2 + \frac{\d r^2}{f_\text{RN}(r)}+r^2\d\Omega^{2}_{2} \ , \quad f_\text{RN}(r) = 1-\frac{2m}{r}+\frac{Q^2}{r^2}\ .
\end{equation}
We now perform our ``regularising" procedure, replacing $r\rightarrow\sqrt{r^2+\ell^2}$ in the metric. The new candidate spacetime is described by the following line element
\begin{equation}
    \d s^2 = -f(r)\d t^2 + \frac{\d r^2}{f(r)} + \left(r^2+\ell^2\right)\d\Omega^{2}_{2} \ , \quad f(r) = 1-\frac{2m}{\sqrt{r^2+\ell^2}}+\frac{Q^2}{r^2+\ell^2} \ .
\end{equation}
One can immediately see that the natural domains for the angular and temporal coordinates are unaffected by the regularisation procedure.
In contrast the natural domain of the $r$ coordinate expands from $r\in[0,+\infty)$ to $r\in(-\infty,+\infty)$. 
Asymptotic flatness is preserved, as are the manifest spherical and time translation symmetries. 
Given the diagonal metric environment, it is trivial to establish the following tetrad
\begin{align}
    \big(e_{\hat{t}}\big)^\mu &= \frac{1}{\sqrt{|f(r)|}} \left( 1, 0, 0, 0 \right), \quad
    \big(e_{\hat{r}}\big)^\mu = \sqrt{|f(r)|} \left(0,1,0,0 \right), \nonumber \\
    \big(e_{\hat{\theta}}\big)^\mu &= \frac{1}{\sqrt{r^2+\ell^2}}\left(0,0,1,0 \right), \quad
    \big(e_{\hat{\phi}}\big)^\mu = \frac{1}{\sqrt{r^2+\ell^2}\,\sin\theta}\left(0,0,0,1 \right),\label{tetrad_bbRN}
\end{align}
and straightforward to verify that this is indeed a solution of $g_{\mu\nu}e_{\hat{\mu}}{}^{\mu}e_{\hat{\nu}}{}^{\nu}=\eta_{\hat{\mu}\hat{\nu}}$. (Note however that the $-1$ in the Minkowski metric corresponds to the timelike direction, and is therefore in the $\hat t \hat t$ position when $f(r)>0$, and in the $\hat r \hat r$ position when instead $f(r)<0$). The analysis that follows is, when appropriate, performed with respect to this orthonormal basis.

Furthermore, note that as $\ell\to0$ one recovers the standard Reissner--Nordstr\"om geometry, while when $m\to0$ and $Q\to0$ one recovers the standard Morris--Thorne wormhole~\cite{Morris-Thorne,MTY,Visser-book,Boonserm:2018}:
\begin{equation}
    \d s^2 = -\d t^2 + \d r^2 + (r^2+\ell^2)\,\d\Omega^{2}_{2} \ .
\end{equation}

\subsection{Kretschmann scalar}\label{SS:Kretschmann}
Our first task consists in showing that our spacetime is indeed everywhere regular. Conveniently, given that our spacetime is static, examination of the Kretschmann scalar $K=R^{\mu\nu\rho\sigma}\,R_{\mu\nu\rho\sigma}$ will be sufficient to accomplish this task~\cite{Lobo:2020}.

Indeed, a simple computation shows that the Kretschmann scalar is quartic in $Q$ and given by
\begin{align}
    K &= \frac{4}{(r^2+\ell^2)^6}\Bigg\lbrace\left(3\ell^4-10\ell^2r^2+14r^4\right)Q^4  \\
    & +\sqrt{r^2+\ell^2}\left[2\ell^2(2\ell^2-3r^2)\sqrt{r^2+\ell^2}-2m(5\ell^4-11\ell^2r^2+12r^4)\right]Q^2 \nonumber \\
    & +3\ell^8+(9m^2+6r^2)\ell^6+3r^2(r^2-m^2)\ell^4-8m\ell^2(\ell^4-r^4)\sqrt{r^2+\ell^2}+ 12m^2r^6\Bigg\rbrace \ . \nonumber
\end{align}
In the limit as $r\rightarrow 0$ we have the finite result
\begin{equation}
    \lim_{r\rightarrow 0} K = \frac{12}{\ell^8}\,Q^4+\frac{8(2\ell-5m)}{\ell^7}\,Q^2+\frac{4\left(3\ell^2-8m\ell+9m^2\right)}{\ell^6} \ .
\end{equation}

So in this manifestly static situation we are then guaranteed that all of the orthonormal curvature tensor components are automatically finite~\cite{Lobo:2020} (nonetheless for completeness we provide  in Appendix \ref{S:invariants} expressions for other curvature invariants).  We may then conclude that the geometry is indeed globally regular.

\subsection{Curvature tensors}\label{S:tensors}

In order to complete our investigation on curvature, we display here the explicit forms for the various non-zero curvature components. 
This is mostly conveniently done in the above introduced orthonormal basis.
The orthonormal components of the Riemann curvature tensor are given by 
\begin{subequations}
\begin{align}
    R^{\hat{t}\hat{r}}{}_{\hat{t}\hat{r}} &= \frac{\ell^2-3r^2}{(r^2+\ell^2)^3}Q^2-\frac{m(\ell^2-2r^2)}{(r^2+\ell^2)^{5/2}}\ ,\\
    R^{\hat{t}\hat{\theta}}{}_{\hat{t}\hat{\theta}} = R^{\hat{t}\hat{\phi}}{}_{\hat{t}\hat{\phi}} &= \frac{r^2}{(r^2+\ell^2)^3}Q^2-\frac{mr^2}{(r^2+\ell^2)^{5/2}}\ ,\\
    R^{\hat{r}\hat{\theta}}{}
    _{\hat{r}\hat{\theta}} = R^{\hat{r}\hat{\phi}}{}_{\hat{r}\hat{\phi}} &= \frac{r^2-\ell^2}{(r^2+\ell^2)^3}Q^2+\frac{m(2\ell^2-r^2)-\ell^2\sqrt{r^2+\ell^2}}{(r^2+\ell^2)^{5/2}}\ ,\\
    R^{\hat{\theta}\hat{\phi}}{}_{\hat{\theta}\hat{\phi}} &= -\frac{r^2}{(r^2+\ell^2)^3}Q^2+\frac{2mr^2+\ell^2\sqrt{r^2+\ell^2}}{(r^2+\ell^2)^{5/2}} \ ,
\end{align}
\end{subequations}
and in the limit as $r\rightarrow 0$ we find
\begin{equation}
    R^{\hat{t}\hat{r}}{}_{\hat{t}\hat{r}}\rightarrow \frac{Q^2-m\ell}{\ell^4} \ , \quad R^{\hat{t}\hat{\theta}}{}_{\hat{t}\hat{\theta}}\rightarrow 0 \ , \quad R^{\hat{r}\hat{\theta}}{}_{\hat{r}\hat{\theta}}\rightarrow -\frac{Q^2+\ell^2-2m\ell}{\ell^4} \ , \quad R^{\hat{\theta}\hat{\phi}}{}_{\hat{\theta}\hat{\phi}}\rightarrow \frac{1}{\ell^2} \ .
\end{equation}
For the orthonormal components of the Ricci tensor we find
\begin{subequations}
\begin{align}
    R^{\hat{t}}{}_{\hat{t}} &= \frac{\ell^2-r^2}{(r^2+\ell^2)^{3}}Q^2 - \frac{m\ell^2}{(r^2+\ell^2)^{5/2}}\ ,\\
    R^{\hat{r}}{}_{\hat{r}} &= -\frac{Q^2}{(r^2+\ell^2)^2} + \frac{\ell^2(3m-2\sqrt{r^2+\ell^2})}{(r^2+\ell^2)^{5/2}}\ ,\\
    R^{\hat{\theta}}{}_{\hat{\theta}} &= R^{\hat{\phi}}{}_{\hat{\phi}} 
    = \frac{r^2-\ell^2}{(r^2+\ell^2)^3}Q^2+\frac{2m\ell^2}{(r^2+\ell^2)^{5/2}} \ ,
\end{align}
\end{subequations}
and in the limit as $r\rightarrow 0$
\begin{equation}
    R^{\hat{t}}{}_{\hat{t}}\rightarrow \frac{Q^2-m\ell}{\ell^4} \ ; \quad R^{\hat{r}}{}_{\hat{r}}\rightarrow \frac{3m\ell-2\ell^2-Q^2}{\ell^4} \ ; \quad R^{\hat{\theta}}{}_{\hat{\theta}}= R^{\hat{\phi}}{}_{\hat{\phi}}\rightarrow \frac{2m\ell-Q^2}{\ell^4} \ .
\end{equation}
Finally, for the Weyl tensor we obtain
\begin{align}
    C^{\hat{t}\hat{r}}{}_{\hat{t}\hat{r}} &= -2C^{\hat{t}\hat{\theta}}{}_{\hat{t}\hat{\theta}} = -2C^{\hat{t}\hat{\phi}}{}_{\hat{t}\hat{\phi}} = -2C^{\hat{r}\hat{\theta}}{}_{\hat{r}\hat{\theta}} = -2C^{\hat{r}\hat{\phi}}{}_{\hat{r}\hat{\phi}} = C^{\hat{\theta}\hat{\phi}}{}_{\hat{\theta}\hat{\phi}} \nonumber \\
    &= \frac{2(\ell^2-3r^2)}{3(r^2+\ell^2)^3}Q^2+\frac{2\ell^2\sqrt{r^2+\ell^2}-3m(\ell^2-2r^2)}{3(r^2+\ell^2)^{5/2}} \ ,
\end{align}
and in the limit as $r\rightarrow 0$
\begin{equation}
    C^{\hat{t}\hat{r}}{}_{\hat{t}\hat{r}}\rightarrow \frac{2Q^2+2\ell^2-3m\ell}{3\ell^4} \ .
\end{equation}

Now that the regularity of spacetime is out of question, we shall proceed by discussing its geometrical properties, \emph{i.e.}\ horizons and characteristic orbits.
\subsection{Horizons and surface gravity}\label{horizon}
In view of the diagonal metric environment, horizon locations are characterised by
\begin{align}
    g_{tt} &=
    0\quad \Longrightarrow \quad r_{H} = S_1 \sqrt{ \left(m+S_2\sqrt{m^2-Q^2}\right)^2 -\ell^2 } \ .
\end{align}
Here $S_{1},S_{2}=\pm 1$, and choice of sign for $S_{1}$ dictates which universe we are in, whilst the choice of sign on $S_{2}$ corresponds to an outer/inner horizon respectively.
For horizons to exist we need both $|Q|\leq m$ and $\ell \leq m \pm \sqrt{m^2-Q^2}$.
The case $|Q|=m$ while $\ell \leq m $ leads to extremal horizons at
$r_H = \pm \sqrt{m^2-\ell^2}$. When $|Q|>m$ there are no horizons, and the geometry is that of a traversable wormhole. Finally when $|Q|\leq m$ but $\ell$ is large enough, $\ell > m \pm \sqrt{m^2-Q^2}$, first the inner horizons vanish and then the outer horizons vanish.

\begin{figure}
    \centering
    \includegraphics[width=.65\textwidth]{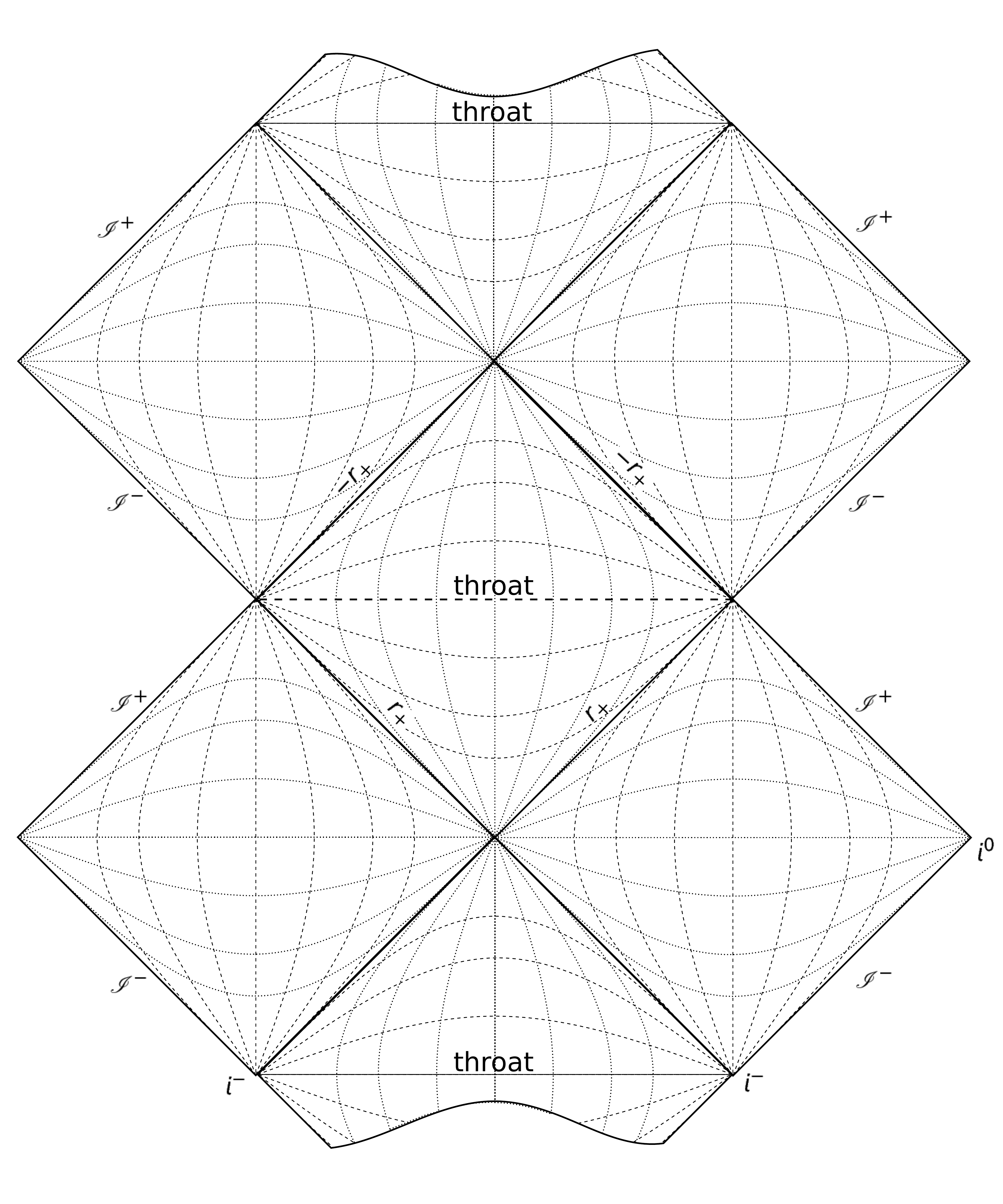}
    \caption{Penrose diagram for a regular black hole with only outer horizons, corresponding to $m-\sqrt{m^2-Q^2} < \ell < m+\sqrt{m^2-Q^2}$. The lower (upper) portion of the diagram corresponds to the $r>0$ ($r<0$) universe; the diagram continues indefinitely above and below the portion shown by repetition of this fundamental block. Here, $ r_+$ is $r_H$ with $S_2=+1$; the sign in front of it is $S_1$.\label{fig:penrose1}}
\end{figure}

\begin{figure}
    \centering
    \includegraphics[width=.9\textwidth]{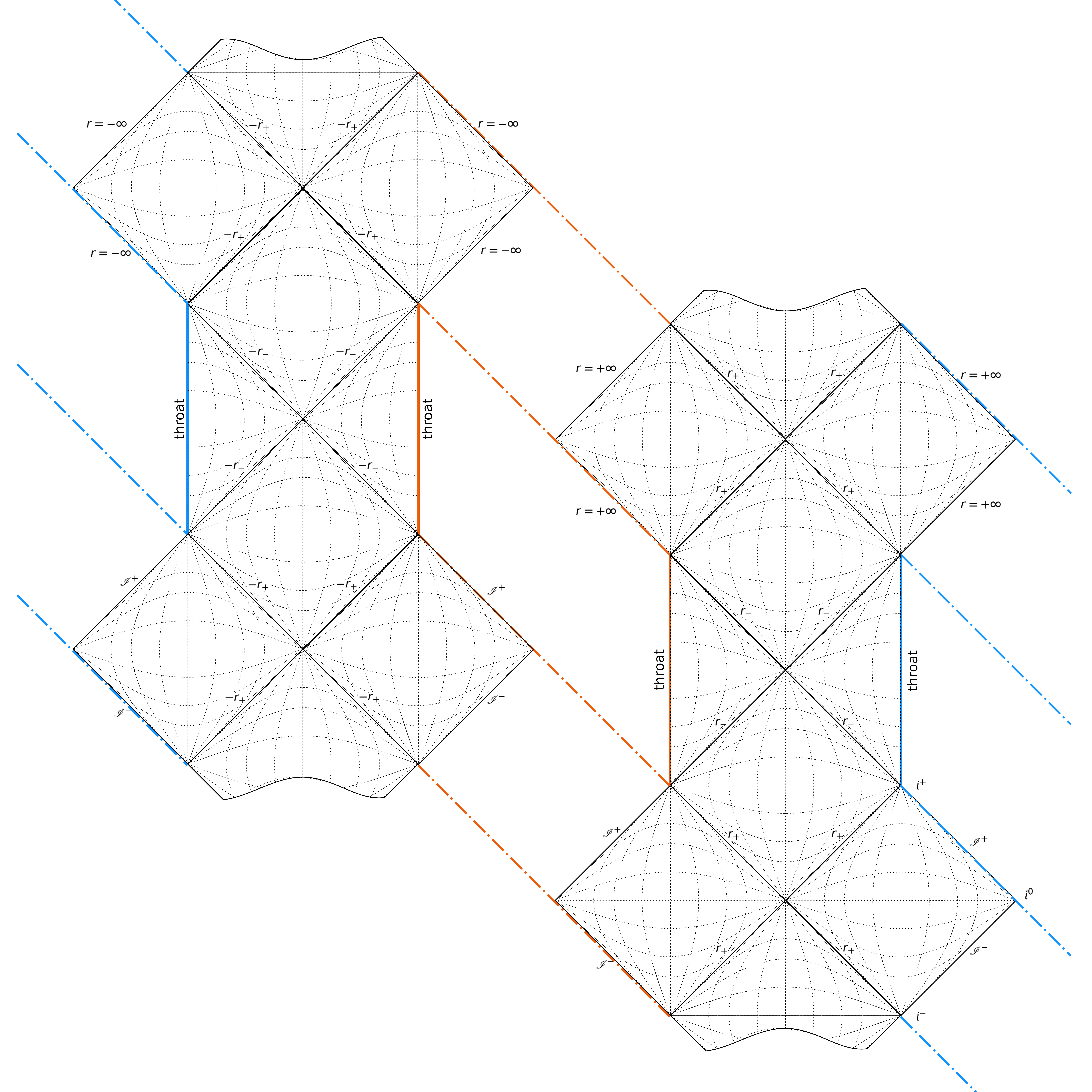}
    \caption{Penrose diagram for a regular black hole with outer and inner horizons, corresponding to $\ell < m-\sqrt{m^2-Q^2}$. Vertical lines of the same colour are identified, as the right-hand (left-hand) part of the diagram represents the $r>0$ ($r<0$) universe; the diagram continues indefinitely above and below the portion shown. Here $r_+$ (resp.~$r_-$) is $r_H$ with $S_2=+1$ ($-1$); the sign in front of it is $S_1$.\label{fig:penrose2}}
\end{figure}

The structure of the maximally-extended spacetime can be visualised with the aid of the Penrose diagrams in figures \ref{fig:penrose1} and \ref{fig:penrose2} for the two qualitatively different regular black hole cases. Both diagrams will be relevant for the rotating generalisation of the next section, too; they are analogous to those of \cite{Liberati:2021} (both) and \cite{Simpson:2018} (the first one).

It is straightforward to calculate the surface gravity at the event horizon in \emph{our} universe for the black-bounce--Reissner--Nordstr\"om spacetime. Keeping in mind that we are still working in curvature coordinates, the surface gravity $\kappa_{H}$ reduces to~\cite{Visser:1992}
\begin{equation}
    \kappa_{H} = \lim_{r\rightarrow r_{H}}\frac{1}{2}\frac{\partial_{r}g_{tt}}{\sqrt{-g_{tt}g_{rr}}} \ .
\end{equation}
For our metric we have $g_{tt} = -1/g_{rr}$ so this simplifies to
\begin{align}
    \kappa_{H} = \left.\frac{1}{2}\partial_{r}g_{tt}\right|_{r_H}
    = \frac{\sqrt{(m+\sqrt{m^2-Q^2})^2-\ell^2} \; \sqrt{m^2-Q^2}}{(m+\sqrt{m^2-Q^2})^3}
    = \kappa_H^\text{RN} \sqrt{\frac{r_H^2}{r_H^2+\ell^2}} \ ,
\end{align}
where $\kappa_H^\text{RN}$ is the surface gravity of a Reissner--Nordstr\"om black hole of the same mass and charge, and as usual the associated Hawking temperature is given by $k_{B}T_{H} = \frac{\hbar}{2\pi}\kappa_{H}$~\cite{Visser:1992}.
(This gives the usual Reissner--Nordstr\"om result as $\ell\to0$.)

\subsection{Innermost stable circular orbit and photon sphere}

Let us briefly examine the coordinate locations of the notable orbits for our candidate spacetime, the innermost stable circular orbit (ISCO) and photon sphere~\cite{Boonserm:isco-osco, Berry:2020,Berry:2021}. Firstly we define the object
\begin{equation}
\epsilon = 
\begin{cases}
-1 & \quad\mbox{massive particle, \emph{i.e.}\ timelike worldline} \\
 0 & \quad\mbox{massless particle, \emph{i.e.}\ null worldline}.
\end{cases}
\end{equation}
Considering the affinely parameterised tangent vector to the worldline of a massive or massless particle, and fixing $\theta=\pi/2$ in view of spherical symmetry, we obtain the reduced equatorial problem
\begin{equation}
    \frac{\d s^2}{\d\lambda^2} = -g_{tt}\left(\frac{\d t}{\d\lambda}\right)^2+g_{rr}\left(\frac{\d r}{\d\lambda}\right)^2+(r^2+\ell^2)\left(\frac{\d\phi}{\d\lambda}\right)^2 = \epsilon \ .
\end{equation}
The Killing symmetries yield the following expressions for the conserved energy $E$, and angular momentum per unit mass $L$
\begin{equation}
    \left(1-\frac{2m}{\sqrt{r^2+\ell^2}}+\frac{Q^2}{r^2+\ell^2}\right)\left(\frac{\d t}{\d\lambda}\right) = E \ ; \quad (r^2+\ell^2)\left(\frac{\d\phi}{\d\lambda}\right) = L \ ,
\end{equation}
yielding the ``effective potential'' for geodesic orbits
\begin{equation}
    V_{\epsilon}(r) = \left(1-\frac{2m}{\sqrt{r^2+\ell^2}}+\frac{Q^2}{r^2+\ell^2}\right)\left[-\epsilon+\frac{L^2}{r^2+\ell^2}\right] \ .
\end{equation}

\subsubsection*{Null orbits}
For the massless case of null orbits, \emph{e.g.}\ photon orbits, set $\epsilon=0$ and solve $V_{0}'(r) = 0$ for the location of the ``photon sphere". We have
\begin{align}
    V_{0}'(r) &= \frac{2L^2r\left[3m-\sqrt{r^2+\ell^2}\right]}{(r^2+\ell^2)^{5/2}} - \frac{4L^2r}{(r^2+\ell^2)^{3}}Q^2 \ ,
\end{align}
yielding the analytic location for the photon sphere in \emph{our} universe \emph{outside} horizons:
\begin{equation}
    r_{\gamma} = \sqrt{\frac{m}{2}\left(9m+3\sqrt{9m^2-8Q^2}\right)-2Q^2-\ell^2} \ .
\end{equation}
In the limit as $Q,\ell\rightarrow 0$ we reproduce the standard Schwarzschild result, $r_{\gamma}=3m$, as expected.

\subsubsection*{Timelike orbits}

For the massive case of timelike orbits, the ISCO is found \emph{via} setting $V_{-1}'(r)=0$. We have
\begin{align}
    V_{-1}'(r) &= \frac{2r\left[m(r^2+\ell^2)+L^2(3m-\sqrt{r^2+\ell^2})\right]}{(r^2+\ell^2)^{5/2}} - \frac{2r(2L^2+r^2+\ell^2)}{(r^2+\ell^2)^{3}}Q^2 \ .
\end{align}
Equating $V_{-1}'(r)=0$ and solving for $r$ is not analytically feasible. We may make life easier \emph{via} the change of variables $z=\sqrt{r^2+\ell^2}$, giving
\begin{equation}
    V_{-1}'(z) = -\frac{2\sqrt{z^2-\ell^2}}{z^6}\left\lbrace (2L^2+z^2)Q^2+z\left[L^2(z-3m)-mz^2\right]\right\rbrace \ .
\end{equation}
Assuming some fixed orbit at some $r_{c}$, hence fixing the corresponding $z_{c}=\sqrt{r_{c}^{2}+\ell^2}$, we may rearrange to find the required angular momentum per unit mass $L_{c}$ as a function of $z_{c}$ and the metric parameters. It follows that the ISCO will be located at the coordinate location where $L_{c}$ is minimised~\cite{Boonserm:isco-osco,Berry:2020,Berry:2021}. 
We find
\begin{align}
    L_{c} &= \frac{z_{c}\sqrt{mz_{c}-Q^2}}{\sqrt{z_{c}^{2}-3mz_{c}+2Q^2}} \ ; \quad
    \Longrightarrow \quad \frac{\partial L_{c}}{\partial z_{c}} = -\frac{6m^2z_{c}^2-mz_{c}(9Q^2+z_{c}^2)+4Q^4}{2\sqrt{mz_{c}-Q^2}(z_{c}^2-3mz_{c}+2Q^2)^{3/2}} \ .
\end{align}
Equating $\partial L_{c}/\partial z_{c}=0$ and solving for $z_{c}$, rearranging for $r_{c}$, and discounting complex roots leaves the analytic ISCO location for timelike particles in \emph{our} universe:
\begin{equation}
r_c = \frac{\sqrt{9 m^4 Q^4 - 6 m^2 Q^2 \left(A^2+2 A m^2+4 m^4\right) + \left(A^2+2 A m^2+4 m^4\right)^2 - A^2 \ell^2 m^2}}{mA}
\end{equation}
where now 
\begin{align}
    A &= \left[2m^2Q^4+m^2\left(\pm B-9m^2\right)Q^2+8m^6\right]^{\frac{1}{3}} \ ; \nonumber \\
    B &= \sqrt{4Q^4-9m^2Q^2+5m^4} \ .
\end{align}
It is easily verified that in the limit as $Q,\ell\rightarrow 0$, $r_{c}\rightarrow 6m$, as expected for Schwarzschild.

\subsection{Stress-energy tensor}\label{stress-energy}

Starting from a given geometry, the discussion of the associated stress-energy tensor necessarily requires us to fix the geometrodynamics. In the following we shall assume that this is everywhere described by general relativity. While this might seem a crude assumption for geometries associated to a possible regularisation of singularities by quantum gravity, we do expect it --- once the regular spacetime settles down in an equilibrium state after the collapse --- to be a good approximation everywhere, if the final regularisation scale $\ell$ is much larger than the Planck scale, or at least sufficiently far away from the core region otherwise.

Given the above assumption, the determination of the stress-energy tensor associated to our spacetime is easily accomplished by computing the non-zero components of the Einstein tensor.
This leads to the following decomposition for the total stress-energy tensor $T^{\hat{\mu}}{}_{\hat{\nu}}$ valid \emph{outside} the outer horizon (and \emph{inside} the inner horizon):
\begin{equation}
    \frac{1}{8\pi}\,G^{\hat{\mu}}{}_{\hat{\nu}} = T^{\hat{\mu}}{}_{\hat{\nu}} 
    = \left[T_\text{bb}\right]^{\hat{\mu}}{}_{\hat{\nu}} + \left[T_{Q}\right]^{\hat{\mu}}{}_{\hat{\nu}} 
    = \diag\left(-\varepsilon, p_r, p_t, p_t\right) \ . 
\end{equation}
In contrast, \emph{between} the inner and outer horizons we have:
\begin{equation}
    \frac{1}{8\pi}\,G^{\hat{\mu}}{}_{\hat{\nu}} 
    = T^{\hat{\mu}}{}_{\hat{\nu}} 
    = \left[T_\text{bb}\right]^{\hat{\mu}}{}_{\hat{\nu}} + \left[T_{Q}\right]^{\hat{\mu}}{}_{\hat{\nu}} 
    = \diag\left(p_r, -\varepsilon, p_t, p_t\right) \ . 
\end{equation}
Here $\left[T_\text{bb}\right]^{\hat{\mu}}{}_{\hat{\nu}}$ is the stress-energy tensor for the original electrically neutral black-bounce spacetime from ref.~\cite{Simpson:2018}, and $\left[T_{Q}\right]^{\hat{\mu}}{}_{\hat{\nu}}$ is the charge-dependent contribution to the stress-energy. 
Examination of the non-zero components of the Einstein tensor, outside the outer horizon (and inside the inner horizon), yields the following:
\begin{subequations}
\begin{align}\label{rho}
    \varepsilon &= \frac{Q^2(r^2-2\ell^2)}{8\pi(r^2+\ell^2)^{3}} + \frac{\ell^2\left(4m-\sqrt{r^2+\ell^2}\right)}{8\pi(r^2+\ell^2)^{5/2}} \ ,\\
    -p_{r} &= \frac{Q^2r^2}{8\pi(r^2+\ell^2)^{3}} + \frac{\ell^2}{8\pi(r^2+\ell^2)^{2}} \ ,\\
    p_{t} &= \frac{Q^2r^2}{8\pi(r^2+\ell^2)^{3}} + \frac{\ell^2\left(\sqrt{r^2+\ell^2}-m\right)}{8\pi(r^2+\ell^2)^{5/2}} \ .
\end{align}
\end{subequations}
For the radial null energy condition (NEC)~\cite{Visser-book,Visser:1989a,Visser:1989b, Visser:2003,Kar:2004,gvp1,gvp2,gvp4,EC-survey-LNP,Martin-Moruno:2013,twilight}, outside (inside) the outer (inner) horizon, we have:
\begin{align}
    \varepsilon + p_{r} 
    = -\frac{\ell^2\left[r^2+\ell^2-2m\sqrt{r^2+\ell^2}+Q^2\right]}{4\pi(r^2+\ell^2)^{3}}
    = -\frac{\ell^2 f(r)}{4\pi(r^2+\ell^2)^{2}}\ .
\end{align}
It is then clear that outside the outer horizon (or inside the inner one), where $f(r)>0$, we will have violation of the radial NEC. 
At the horizons, where $f(r)=0$, we always have $\left.\left(\varepsilon+p_r\right)\right|_H = 0$. 
(This on-horizon marginal satisfaction of the NEC is a quite generic phenomenon~\cite{Martin-Moruno:2021,Visser:1992,Medved:2004-1,Medved:2004-2}.)

We can trivially extract the explicit form for $\left[T_{Q}\right]^{\hat{\mu}}{}_{\hat{\nu}}$:
\begin{align}\label{Tem}
    \left[T_{Q}\right]^{\hat{\mu}}{}_{\hat{\nu}} 
    &= \frac{Q^2r^2}{8\pi(r^2+\ell^2)^{3}} \diag\left( \frac{2\ell^2}{r^2}-1, -1, 1, 1\right) \nonumber \\
    &= \frac{Q^2r^2}{8\pi(r^2+\ell^2)^{3}}\left[\diag\left(-1, -1, 1, 1\right)+ \diag\left(\frac{2\ell^2}{r^2}, 0, 0, 0\right)\right] \ .
\end{align}
In the current situation, the first term above can be interpreted as the usual Maxwell stress-energy tensor~\cite{MisnerThorneWheeler1973}
\begin{equation}\label{TMaxwell}
  \left[T_\text{Maxwell}\right]^{\hat{\mu}}{}_{\hat{\nu}} 
  = \frac{1}{4\pi}
  \left[
  -F^{\hat{\mu}}{}_{\hat{\alpha}}F^{\hat{\alpha}}{}_{\hat{\nu}}
  -\frac{1}{4}\delta^{\hat{\mu}}{}_{\hat{\nu}}F^{2}
  \right] \ ,
\end{equation}
while the second term can be interpreted as the stress-energy of ``charged dust'', with the density of the dust involving both the bounce parameter~$\ell$ and the total charge~$Q$. 
Overall we have
\begin{equation}
  \left[T_{Q}\right]^{\hat{\mu}}{}_{\hat{\nu}} 
  = \left[T_\text{Maxwell}\right]^{\hat{\mu}}{}_{\hat{\nu}}
  +\Xi\; V^{\hat{\mu}}V_{\hat{\nu}}\ .
\end{equation}
The vector $V^{\hat\mu}$ is the normalised unit timelike eigenvector of the stress-energy, which in the current situation reduces to the normalised time-translation Killing vector, while the dust density $\Xi$ has to be determined. We obtain
\begin{equation}
    \left[T_{Q}\right]^{\hat{t}}{}_{\hat{t}} 
    = -\varepsilon_\text{em} = \left[T_\text{Maxwell}\right]^{\hat{t}}{}_{\hat{t}}  -\Xi= -\frac{1}{8\pi}E^2-\Xi \ .
\end{equation}
Comparing with eq.~(\ref{Tem}), we find for the electric field strength $E$
\begin{align}\label{eq:ERN}
    E &= \frac{Qr}{(r^2+\ell^2)^{3/2}} = E_\text{RN}\left[\frac{r^3}{(r^2+\ell^2)^{3/2}}\right] \ ,
\end{align}
where $E_\text{RN}$ is the electric field strength of a  Reissner--Nordstr\"om  black hole.
For the density of the dust $\Xi$ we find
\begin{equation}
    \Xi = -\frac{1}{4\pi} \frac{Q^2\ell^2}{(r^2+\ell^2)^{3}} \ .
\end{equation}
All told, we have the following form for the electromagnetic stress-energy tensor for our regularised Reissner--Nordstr\"om spacetime
\begin{equation}
    \left[T_{Q}\right]^{\hat{\mu}}{}_{\hat{\nu}} 
    = \frac{1}{4\pi}\left[-F^{\hat{\mu}}{}_{\hat{\alpha}}F^{\hat{\alpha}}{}_{\hat{\nu}}-\frac{1}{4}\delta^{\hat{\mu}}{}_{\hat{\nu}}F^2\right] 
    -\frac{1}{4\pi}\frac{Q^2\ell^2}{(r^2+\ell^2)^3}V^{\hat{\mu}}V_{\hat{\nu}}\ .
\end{equation}

Finally, the electromagnetic potential is easily extracted \emph{via} integrating eq.~\eqref{eq:ERN}, and in view of asymptotic flatness we may set the constant of integration to zero, yielding
\begin{equation}
    A_{\mu} = (\Phi_\text{em}(r), 0,0,0) = -\frac{Q}{\sqrt{r^2+\ell^2}}\;(1,0,0,0) .
    \label{eq:emstatic}
\end{equation}
Note that this really is simply the electromagnetic potential from standard Reissner--Nordstr\"om spacetime, $-Q/r$, under the map $r\to \sqrt{r^2+\ell^2}$.

It is easy to verify that the electromagnetic field-strength tensor $F_{\mu \nu} = \nabla_{\mu} A_{\nu} - \nabla_{\nu} A_{\mu}$ satisfies $F_{[\mu\nu,\sigma]}=0$. The inhomogeneous Maxwell equation is, using $z=\sqrt{r^2+\ell^2}$: 
\begin{align}
    \nabla^{\hat {\mu}} F_{\hat{\mu} \hat {\nu}} &= {\frac{Q\ell^2}{z^6}\;\sqrt{z^2-2mz+Q^2}}\; \left(1,0,0,0\right).\label{eq:maxinhomoRN}
\end{align}
(We warn the reader that the situation will get somewhat messier once we add rotation.)

\section{Black-bounce--Kerr--Newman geometry}

As expected, the  black-bounce--Kerr--Newman geometry is qualitatively more complicated (but also physically richer) than the black-bounce--Reissner--Nordstr\"om one. 
We start from Kerr--Newman geometry in standard Boyer--Lindquist coordinates~\cite{Kerr-intro,Kerr-book}
\begin{equation}
    \d s^2_\text{KN}  = -\frac{\Delta_\text{KN}}{\rho^2_\text{KN}}(a\sin^2\theta \d\phi - \d t)^2 
    + \frac{\sin^2\theta}{\rho^2_\text{KN}}\left[(r^2+a^2)\d\phi -a\d t\right]^2 
    + \frac{\rho^2_\text{KN}}{\Delta_\text{KN}}\d r^2+\rho^2_\text{KN}\d\theta^2 \ ,
\end{equation}
where
\begin{equation}
    \rho^2_\text{KN} = r^2+a^2\cos^2\theta \ ; \qquad \Delta_\text{KN} = r^2+a^2-2mr+Q^2 \ .
\end{equation}
Applying our procedure, our new candidate spacetime is given by the line element
\begin{equation}
    \d s^2 = -\frac{\Delta}{\rho^2}(a\sin^2\theta \d\phi - \d t)^2 + \frac{\sin^2\theta}{\rho^2}\left[(r^2+\ell^2+a^2)\d\phi -a\d t\right]^2 + \frac{\rho^2}{\Delta}\d r^2+\rho^2\d\theta^2 \ ,
    \label{eq:bbKN}
\end{equation}
where now $\rho^2$ and $\Delta$ are modified:
\begin{equation}
    \rho^2 = r^2+\ell^2+a^2\cos^2\theta \ ; \qquad \Delta = r^2+\ell^2+a^2-2m\sqrt{r^2+\ell^2}+Q^2 \ .
\end{equation}
The natural domains of the angular and temporal coordinates are unaffected,
while the radial coordinate again extends from the positive half line to the entire real line, and both manifest axisymmetry and asymptotic flatness are preserved. This spacetime is now \emph{stationary} but not static, hence we must examine more than just the Kretschmann scalar to draw any conclusion as to regularity. 

\subsection{Curvature tensors}\label{S:tensorsKN}

Again using $z:= \sqrt{r^2 + \ell^2}$, where this shorthand now stands for the equatorial value of the parameter $\rho$, the Ricci scalar is given by
\begin{equation}
R = 2 \ell^2 \frac{m \left(\rho^4-2 z^4\right)+Q^2 z^3+z^3 \left(\rho^2-2 \Delta \right)}{\rho^6 z^3}.
\end{equation}
It is clearly finite in the limit $r\to 0$, \emph{i.e.}\ $z\to \ell$. So are the Kretschmann scalar and the invariants $R^{\mu \nu}R_{\mu \nu}$ and $C^{\mu \nu \rho \sigma}C_{\mu \nu \rho \sigma}$. The only potentially dangerous behaviour arises from the denominators, which, in the $r\to 0$ limit, take the form
\begin{equation}
    \frac{1}{\ell^2 (\ell^2 + a^2 \cos^2\theta)^6}.
\end{equation}
As long as $\ell \neq 0$, therefore, these quantities are never infinite. 

We now turn our attention to the Einstein and Ricci tensors. We note that the (mixed) Einstein tensor $G^\mu _{\ \nu}$ can be diagonalised over the real numbers: its four eigenvectors $\{ e_{\hat{\mu}}\}_{\mu = t,r,\theta,\phi}$ form a globally defined tetrad~\cite{Rajan:tetrad} and have explicit Boyer--Lindquist components
\begin{align}
    \big(e_{\hat{t}}\big)^\mu &= \frac{1}{\sqrt{\rho^2 \abs{\Delta}}} \left( r^2+\ell^2+a^2, 0, 0, a \right), \quad
    \big(e_{\hat{r}}\big)^\mu = \sqrt{\frac{\abs{\Delta}}{\rho^2}} \left(0,1,0,0 \right), \nonumber \\
    \big(e_{\hat{\theta}}\big)^\mu &= \frac{1}{\sqrt{\rho^2}}\left(0,0,1,0 \right), \quad
    \big(e_{\hat{\phi}}\big)^\mu = \frac{1}{ \sin \theta\sqrt{\rho^2}}\left(a \sin^2 \theta,0,0,1 \right).
    \label{eq:tetrad}
\end{align}
Eigenvectors are defined up to multiplicative, possibly dimensionful constants. This choice of normalisation ensures that the tetrad $\{e_{\hat{\mu}}\}$ is orthonormal and reduces to eq.~\eqref{tetrad_bbRN} in the limit $a\to 0$.
We will therefore use the tetrad \eqref{eq:tetrad}, along with the coordinate basis, to express components of tensors.

In particular, the components of the Einstein tensor are
\begin{subequations}
\begin{align}
    G_{\hat t \hat t} &= \sign\Delta \frac{  \ell^2 \left[2 m \left(z^2-\rho^2\right)+z \left(\rho^2-2 \Delta \right)\right]+ Q^2 z \left(\rho^2-\ell^2\right)}{\rho^6 z}\ ,\\
    G_{\hat r \hat r} &= - \sign\Delta \frac{ \ell^2 \left[2 m \left(z^2-\rho^2\right)+\rho^2 z\right]+Q^2 z \left(\rho^2-\ell^2\right) }{\rho^6 z}\ ,\\
    G_{\hat \theta \hat \theta} &= \frac{\ell^2 \left[m \left(-\rho^4-2 \rho^2 z^2+2 z^4\right)+\rho^2 z^3\right]+Q^2 z^3 \left(\rho^2-\ell^2\right)}{\rho^6 z^3}\ ,\\
    G_{\hat \phi \hat \phi} &= \frac{\ell^2 \left[m \left(-\rho^4-2 \rho^2 z^2+6 z^4\right)+z^3 \left(2 \Delta -\rho^2\right)\right]+Q^2 z^3 \left(\rho^2-3 \ell^2\right)}{\rho^6 z^3}\ .
\end{align}
\end{subequations}
We note that
\begin{equation}
  G_{\hat t \hat t} +  
  G_{\hat r \hat r} 
  = -\sign\Delta\; \frac{2\ell^2\Delta}{\rho^6} 
\end{equation}
which is well-behaved at $\Delta=0$.

The Ricci tensor is clearly diagonal, too:
\begin{subequations}
\begin{align}
    R_{\hat t \hat t} &= \sign\Delta \frac{m \ell^2 \left(-\rho^4-2 \rho^2 z^2+4 z^4\right)+Q^2 z^3 \left(\rho^2-2 \ell^2\right)}{\rho^6 z^3}\ ,\\
    R_{\hat r \hat r} &= \sign\Delta \frac{ \ell^2 \left(m \left(\rho^4+2 \rho^2 z^2-4 z^4\right)-2 \Delta  z^3\right)+Q^2 z^3 \left(2 \ell^2-\rho^2\right) }{\rho^6 z^3}\ ,\\
    R_{\hat \theta \hat \theta} &= \frac{Q^2}{\rho^4} - \frac{2 \ell^2}{\rho^6} \left[\Delta +\frac{\rho^2 (m-z)}{z}\right],\\
    R_{\hat \phi \hat \phi} &= \frac{2 m \ell^2 \left(2 z^2-\rho^2\right)+Q^2 z \left(\rho^2-2 \ell^2\right)}{\rho^6 z}\ .
\end{align}
\end{subequations}
Similarly, we note that
\begin{equation}
  R_{\hat t \hat t} +  
  R_{\hat r \hat r} 
  = -\sign\Delta\; \frac{2\ell^2\Delta}{\rho^6} 
\end{equation}
which is well-behaved at $\Delta=0$.

From these expressions one immediately notices that the curvature tensors are rational polynomials in the variable $z = \sqrt{r^2+\ell^2}$, which is strictly positive, and their denominators never vanish. The same is true for the Riemann and Weyl tensors; we thus conclude that the spacetime is free of curvature singularities.

\subsection{Horizons, surface gravity and ergosurfaces}
Horizons are now associated to the roots of $\Delta$:
\begin{equation}
 r_H = S_1 \sqrt{\left(m+ S_2 \sqrt{m^2-Q^2-a^2} \right)^2-\ell^2},
\end{equation}
where $S_1$ and $S_2$ are defined as in section~\ref{horizon}. The spacetime structures corresponding to $m-\sqrt{m^2-a^2-Q^2}<\ell<m+\sqrt{m^2-a^2-Q^2}$ and $\ell< m-\sqrt{m^2-a^2-Q^2}$ are analogous to their non-spinning counterparts and are represented by the diagrams in figures \ref{fig:penrose1} and \ref{fig:penrose2}.

In the Kerr--Newman geometry, one demands $Q^2+a^2 \leq m^2$ to avoid the possibility of naked singularities. In our case, we need not worry about this eventuality and may consider arbitrary values of spin and charge. Thus, if $Q^2+a^2 > m^2$ or $\ell > m +S_2\sqrt{m^2-Q^2-a^2}$, the spacetime has no horizon.

If horizons are present, their surface gravity is given by
\begin{equation}
    \kappa_{S_2} := \frac{1}{2} \dv{}{r} \eval{\bigg( \frac{\Delta}{r^2+\ell^2 + a^2} \bigg)}_{r_H} = \kappa^\text{KN}_{S_2} \sqrt{\frac{r_H^2}{r_H^2+\ell^2}},
\end{equation}
where $\kappa^\text{KN}_{S_2}$ is the surface gravity relative to the inner, when $S_2=-1$, or outer, when $S_2=+1$, horizon of a Kerr--Newman black hole with mass $m$, spin $a$ and charge $Q$.

The ergosurface is determined by $g_{tt}=0$, which is a quadratic equation in $r$. The roots are given by:
\begin{equation}
    r_\text{erg}=S_1 \sqrt{\left(m +S_2 \sqrt{m^2-Q^2-a^2\cos^2\theta} \right)^2-\ell^2},
\end{equation}
where $S_1, S_2$ are as before.

\subsection{Geodesics and equatorial orbits}

Consider a test particle with mass $\mu$, energy $E$, component of angular momentum (per unit mass) along the rotation axis $L_z$ and zero electric charge. Its trajectory $x^\mu(\tau)$ is governed by the following set of first-order differential equations (see \emph{e.g.}\ ref.~\cite{frolov_black_1998}):
\begin{subequations}\begin{align}
    \rho^2 \dv{t}{\tau} &= a(L_z-a E\sin^2 \theta) + \frac{(r^2+\ell^2) + a^2}{\Delta}[E(r^2 + \ell^2 + a^2) - L_z a],\\
    \rho^2 \dv{r}{\tau} &= \pm \sqrt{\mathcal{R}},\label{eq:eomrad}\\
    \rho^2 \dv{\theta}{\tau} &= \pm \sqrt{\Theta},\\
    \rho^2 \dv{\phi}{\tau} &= \frac{L_z}{\sin^2 \theta} - a E + \frac{a}{\Delta} [E(r^2+ \ell^2 + a^2) - L_z a],
\end{align}\end{subequations}
where 
\begin{align}
    \mathcal{R} &= [E (r^2 + \ell^2 + a^2) -L_z a]^2 - \Delta[\mu^2 (r^2+\ell^2) + (L_z -a E)^2 + \mathcal{Q}],\label{eq:eom1}\\
    \Theta &= \mathcal{Q} - \cos^2 \theta \bigg[ a^2 (\mu^2-E^2) + \frac{L_z^2}{\sin^2 \theta} \bigg],
    \label{eq:eom2}
\end{align}
and $\mathcal{Q}$ is a generalised Carter constant associated to the existence of a Killing tensor discussed in section~\ref{s:Killing} below.

In view of the existence of the Killing tensor, there exist orbits that lie entirely on the equatorial plane $\theta=\pi/2$. Exploiting the conserved quantities, their motion is effectively one-dimensional and governed by the effective potential $\mathcal{R}$: circular orbits, in particular, are given by
\begin{equation}
    \mathcal{R} =0 \qquad \text{and} \qquad \dv{\mathcal{R} }{r}=0;
    \label{eq:circ}
\end{equation}
when, in addition, 
\begin{equation}
    \dv[2]{\mathcal{R}}{r}>0,
\end{equation}
the orbits are stable. 

Solutions to \eqref{eq:circ} can be easily found by exploiting known results on the Kerr--Newman geometry~\cite{Kerr-book,Kerr-intro,Newman:1965,Carter:1968rr}. Indeed, writing \eqref{eq:eom1} in terms of $z:= \sqrt{r^2+\ell^2}$, one immediately recognises the textbook result for a Kerr--Newman spacetime in which the Boyer--Lindquist radius has been given the uncommon name $z$. Moreover, 
\begin{equation}
    \dv{\mathcal{R}}{r} = \dv{z}{r}\; \dv{\mathcal{R}}{z},
\end{equation}
so 
\begin{equation}
    \dv{\mathcal{R}}{z} = 0 \quad \Longrightarrow \quad \dv{\mathcal{R}}{r} = 0.
\end{equation}
Furthermore, at the critical point
\begin{equation}
    \frac{\d^2\mathcal{R}}{\d r^2} = 
    \left(\frac{\d z}{\d r}\right)^2 \;
    \frac{\d^2\mathcal{R}}{\d z^2}.
\end{equation}
So stability (or lack thereof) is unaffected by the substitution  $r\longleftrightarrow z$.
Therefore, suppose $z_0$ is such that
\begin{equation}
    \mathcal{R}(z_0) = 0 \qquad \text{and} \qquad \dv{\mathcal{R} (z_0)}{z} =0;
\end{equation}
that is, suppose the Kerr--Newman spacetime has a circular orbit at radius $z=z_0$, then the black--bounce-Kerr--Newman spacetime has a circular orbit at $r = r_0 := \sqrt{z_0^2-\ell^2}$. Clearly, this mapping is allowed only if $z_0 \geq \ell$. 

Non-circular and non-equatorial orbits, instead, require a more thorough analysis.

\subsection{Killing tensor and non-existence of the Killing tower\label{s:Killing}}

The existence of the generalised Carter constant $\mathcal{Q}$ introduced in the previous section is guaranteed by the fact that the tensor 
\begin{equation}
    K_{\mu \nu} = \rho^2 \left(l_\mu n_\nu + l_\nu n_\mu \right) + \left( r^2 + \ell^2 \right)g_{\mu \nu}
    \label{eq:killingtens}
\end{equation}
is a Killing tensor; it is easy to explicitly check that $K_{(\mu\nu;\lambda)}=0$. Here 
\begin{equation}
    l^\mu = \left(\frac{r^2+\ell^2+a^2}{\Delta}, 1, 0, \frac{a}{\Delta}\right)
    \quad \text{and} \quad
    n^\mu = \frac{1}{2\rho^2}\left(r^2+\ell^2+a^2, - \Delta, 0, a \right)
\end{equation}
are a pair of geodesic null vectors belonging to a generalised Kinnersley tetrad --- see ref.~\cite{Liberati:2021}.

Based on proposition~1.3 in~\cite{Giorgi:2021skz}, it has recently been established~\cite{Baines:2021prep} that when one defines the Carter operator $\mathcal{K}\Phi = \nabla_\mu\left(K^{\mu\nu}\nabla_\nu\Phi\right)$ and wave operator $\Box\Phi = \nabla_\mu\left(g^{\mu\nu}\nabla_\nu\Phi\right)$ one has
\begin{equation}
    \left[\mathcal{K},\Box\right]\Phi = \frac{2}{3}\left(\nabla_\mu\left[R,K\right]^\mu{}_\nu\right)\nabla^\nu\Phi\ .
\end{equation}

This operator commutator will certainly vanish when the tensor commutator $\left[R,K\right]^\mu{}_\nu := R^\mu{}_\alpha K^\alpha{}_\nu - K^\mu{}_\alpha R^\alpha{}_\nu$ vanishes, and this tensor commutator certainly vanishes for the black-bounce--Kerr--Newman spacetime considered herein. Hence the wave equation (not just the Hamilton--Jacobi equation) separates on the black-bounce--Kerr--Newman spacetime.

In the Kerr--Newman spacetime we started from, the Killing tensor is part of a ``Killing tower'' which ultimately descends from the existence of a closed conformal Killing--Yano tensor;  called a \emph{principal tensor} for short~\cite{frolov_black_2017}. Such a principal tensor is a rank-2, antisymmetric tensor $h_{\mu \nu}$ satisfying (in four spacetime dimensions) the equation:
\begin{equation}
    \nabla_\mu h_{\nu \alpha} = \frac{1}{3} \left[ g_{\mu \nu} \nabla^\beta h_{\beta \alpha} - g_{\mu \alpha} \nabla^\beta h_{\beta \nu} \right].
    \label{eq:printenseq}
\end{equation}
In the language of forms, $\mathbf{h}$ is a non-degenerate two-form satisfying 
\begin{equation}
    \mathbf{\nabla_Y h} = \mathbf{Y} \wedge \mathbf{X}, \qquad \mathbf{X} = \frac{1}{3} \div{\mathbf{h}} 
\end{equation}
with $\mathbf{Y}$ any vector. (The equation above implies, incidentally, that $\mathbf{h}$ is closed:
$\mathbf{\dd b}=0 $ so that locally
$\mathbf{h} = \mathbf{\dd b} $.) 
The Hodge dual of a principal tensor is a Killing--Yano tensor, \emph{i.e.}
\begin{align}
    \mathbf{f} = \mathbf{\ast h } \qquad \text{is such that} \qquad
    \nabla_\mu f_{\nu \alpha} + \nabla_\nu f_{\mu \alpha} = 0.
    \label{eq:KYeq}
\end{align}
A Killing--Yano tensor, in turn, squares to a tensor
\begin{equation}
    k_{\mu \nu}:= f_{\mu \alpha} f_\nu^{\ \alpha}
\end{equation}
that is a Killing tensor; $ k_{(\mu \nu;\lambda)}=0 $.

We may thus wonder whether the Killing tensor \eqref{eq:killingtens} derives from a principal tensor, as in the Kerr--Newman case. Naively, one may want to apply the usual trick $r \to \sqrt{r^2+\ell^2}$ to the Kerr--Newman principal tensor, or to the potential $\mathbf{b}$ (the two options are not equivalent). By adopting the first strategy, one finds a ``would-be'' Killing--Yano tensor that does indeed square to \eqref{eq:killingtens} but fails to satisfy eq.~\eqref{eq:KYeq}. The second approach also fails.

In fact, one can prove that no principal tensor can exist in this spacetime. The system \eqref{eq:printenseq} is overdetermined and has a solution only if a certain integrability condition is satisfied: This condition implies that the corresponding spacetime be of Petrov type D. However, three of the current authors proved, in ref.~\cite{Liberati:2021}, that the black-bounce--Kerr spacetime is not algebraically special, hence neither can the black-bounce--Kerr--Newman 
spacetime be algebraically special.
More prosaically, the non-existence of the Killing tower can be seen as a side effect of the fact that our black-bounce--Kerr--Newman geometry does not fall into Carter's ``off shell'' 2-free-function distortion of Kerr~\cite{frolov_black_2017}. 

For reference, here is the would-be Killing--Yano tensor:
\begin{align}
    f_{\mu \nu} &= \begin{pmatrix}
    0 & -a \cos \theta & 0 & 0 \\
    a \cos \theta & 0 & 0 & - a^2 \cos \theta \sin^2 \theta\\
    0& 0 & 0 & 0\\
    0 & a^2 \cos \theta \sin^2 \theta & 0 & 0
    \end{pmatrix}
\nonumber\\
    &+ \sqrt{r^2+\ell^2} \sin\theta \begin{pmatrix}
    0 & 0 & a  & 0 \\
    0 & 0 & 0 & 0\\
    - a & 0 & 0 & (r^2+\ell^2+a^2)\\
    0 & 0 & - (r^2+\ell^2+a^2) & 0
    \end{pmatrix}.
\end{align}
This would-be Killing--Yano tensor is taken from~\cite[eq.~(3.22), p.~47] {frolov_black_2017}, with coordinates changed to Boyer--Lindquist form, and with the substitution $r\to \sqrt{r^2+\ell^2}$ in the tensor components. 
It is easy to check that $ f^2 = K$, but 
\begin{equation}
\nabla_{(\mu} f_{\nu) \alpha} = \left(\sqrt{r^2+\ell^2}-r \right) \times \left[\text{tensor  that is finite as }\ell \to 0 \right].
\end{equation}
(This manifestly vanishes when $\ell\to0$, as it should to recover the Killing--Yano tensor of the Kerr--Newman spacetime.)
Its divergence is in fact particularly simple:
\begin{equation}
    \nabla_\mu f^{\mu \nu} = 
    \left[\left( \sqrt{r^2+\ell^2}-r \right)\frac{2a \cos \theta}{\rho^2}\right]
    \left(1, 0, 0, 0 \right).
\end{equation}
(This again manifestly vanishes when $\ell\to0$ as it should.)

Note that if one instead takes 
\begin{equation}
    b_\mu \dd x^\mu = -\frac{1}{2} (r^2 +\ell^2-a^2\cos^2\theta) \dd t -\frac{1}{2} \left[-r^2-\ell^2 + (r^2+\ell^2+a^2)\cos^2\theta \right] a \dd \phi,
\end{equation}
as in ref.~\cite[eq.~(3.21), p.~47]{frolov_black_2017}, 
converted to Boyer--Lindquist coordinates, 
and subjected to the substitution $r\to\sqrt{r^2+\ell^2}$, one finds
\begin{equation}
f_{\mu \nu} \neq \nabla_\mu b_\nu - \nabla_\nu b_\mu.
\end{equation}
(This is not surprising since derivatives are involved.)

Having now completed a purely geometrical treatment of the properties of our spacetime we can move on to discuss the implications of choosing a definite geometrodynamics, namely, as before, that of general relativity.

\subsection{Stress-energy tensor}
We may again exploit the orthonormal tetrad \eqref{eq:tetrad} to characterise the distribution of stress-energy in our spacetime. Assuming standard general relativity holds, the Einstein tensor is proportional to the stress-energy tensor: we may thus interpret the one component of $G_{\hat{\mu} \hat{\nu}}$ that corresponds to the timelike direction as an energy density $\varepsilon$, and all the other non-zero components as principal pressures $p_i$. 

In particular, outside any horizon (technically, whenever $\Delta > 0$), we have
\begin{subequations}
\begin{align}
    \varepsilon &= -\frac{\ell^2 \left(2 a^2 z+2 m \rho^2-6 m z^2+2 z^3-\rho^2 z\right)}{8\pi \rho^6 z}-\frac{Q^2 \left(3 \ell^2-\rho^2\right)}{8\pi \rho^6},\\
    p_1 &= \frac{\ell^2 \left[\rho^2 (2 m-z)-2 m z^2\right]}{8\pi \rho^6 z}+\frac{Q^2 \left(\ell^2-\rho^2\right)}{8\pi \rho^6} ,\\
    p_2 &= \frac{\ell^2 \left[-m \rho^4+\rho^2 z^2 (z-2 m)+2 m z^4\right]}{8\pi \rho^6 z^3}+\frac{Q^2 \left(\rho^2-\ell^2\right)}{8\pi \rho^6} ,\\
    p_3 &= \frac{\ell^2 \left[2 a^2 z^3+m \left(-\rho^4-2 \rho^2 z^2+2 z^4\right)-\rho^2 z^3+2 z^5\right]}{8\pi \rho^6 z^3}+\frac{Q^2 \left(\rho^2-\ell^2\right)}{8\pi \rho^6}.
\end{align}
\end{subequations}
The expressions above prove, incidentally, that our black-bounce--Kerr--Newman spacetime is Hawking--Ellis type I~\cite{Hawking:1973uf,Martin-Moruno:2021,Martin-Moruno:core, Martin-Moruno:Rainich}.

Note that 
\begin{equation}
    \varepsilon + p_1 = - \frac{\ell^2 \Delta}{{8\pi}\rho^6}.
\end{equation}
This is the same result one gets in the black-bounce--Kerr spacetime, modulo the redefinition of $\Delta$. Thus, in particular, the NEC is violated.
Note that on the horizon $\Delta=0 $ so $\left.(\varepsilon+p_1)\right|_H =0$. 
This on-horizon simplification is a useful consistency check~\cite{Martin-Moruno:2021, Medved:2004-1, Medved:2004-2, Visser:1992}. 

We now wish to characterise the spacetime by means of some variant of curved-spacetime Maxwell-like electromagnetism, that is, by assuming that some variant of Maxwell's equations hold. By doing so, as in the non-rotating case, we find that the matter content is made up of two different components: one electrically neutral and one charged, with the charged component further subdividing into Maxwell-like and charged dust contributions. 
If we isolate the $Q$-dependent contribution to the total stress-energy, we find
\begin{equation}
\label{E:xxx}
  [T_Q]^{\hat\mu}{}_{\hat\nu}  = \frac{1}{8\pi} \frac{Q^2 \left(\rho^2-\ell^2\right)}{\rho^6}
  \left[ \diag\left(-1,-1,1,1\right) +\frac{2\ell^2}{\rho^2{-\ell^2}} \diag\left(1,0,0,0\right)
  \right].
\end{equation}

This is structurally the same as what we saw happening for the black-bounce--Reissner--Nordstr\"om spacetime, cfr.\ eq.~\eqref{Tem}, with the substitutions
\begin{equation}
    \frac{Q^2 r^2} {(r^2+\ell^2)^3}
    \longleftrightarrow
    \frac{Q^2 \left(\rho^2-\ell^2\right)}{\rho^6}
    \qquad \hbox{and} \qquad
    \frac{2\ell^2}{r^2}
    \longleftrightarrow
    \frac{2\ell^2}{\rho^2{-\ell^2}}.
\end{equation}
The first term in eq.~\eqref{E:xxx} is structurally of the form of the Maxwell stress-energy tensor, and the second term is structurally of the form of charged dust. At first sight this seems to suggest that a similar treatment as the one presented for the black-bounce--Reissner--Nordstr\"om spacetime should lead to a consistent picture. 
However, we shall soon see below, that this case is quite trickier than the previous one.
%
\subsubsection{Electromagnetic potential and field-strength tensor}
The first step in carrying on the same interpretation for the stress-energy tensor as that applied in the black-bounce--Reissner--Nordstr\"om spacetime, is to introduce the electromagnetic potential. 
Of course, also in this case there is no obvious way to derive it, since we are not \emph{a priori} specifying the equations of motion for the electromagnetic sector. Therefore, we shall choose to modify the Kerr--Newman potential in a minimal way as we did before for the Reissner--Nordstr\"om case, \emph{i.e.}\ by performing the usual substitution $r\to \sqrt{r^2+\ell^2}$. Thus, our proposal in Boyer--Lindquist coordinates reads
\begin{equation}
    A_{\mu} = -\frac{Q \sqrt{r^2+\ell^2}}{\rho^2} \left(1,0,0,-a \sin^2\theta \right).
    \label{eq:empot}
\end{equation}
In the orthonormal basis one has
\begin{equation}
    A_{\hat \mu} = e_{\hat \mu}{}^\nu \; A_\nu = 
    -\frac{Q \sqrt{r^2+\ell^2}} {\sqrt{\rho^2|\Delta|}} \left( 1,0,0,0 \right).
    \label{eq:empot2}
\end{equation}
This is a minimal modification in the sense that, when we put $a\to0$, the corresponding electrostatic potential is that of the black-bounce--Reissner--Nordstr\"om spacetime \eqref{eq:emstatic}, and when $\ell\to 0$ we regain the usual result for standard Kerr--Newman.
The potential \eqref{eq:empot} is also compatible with the Newman--Janis procedure as outlined in ref.~\cite{erbin_janis-newman_2017}, and as applied to the
black-bounce--Reissner--Nordstr\"om geometry.

We can now compute the electromagnetic field-strength tensor
$F_{\mu \nu}$.
In the orthonormal basis, its only non-zero components are
\begin{align}
   F_{\hat t \hat r} = -F_{\hat r \hat t} &=  - \frac{Q}{\rho^4} \sqrt{\frac{r^2 }{r^2+\ell^2}} (r^2 +\ell^2 - a^2 \cos^2\theta)\ ,\\
   F_{\hat \theta \hat \phi} = -F_{\hat \phi \hat \theta} &=  \frac{2aQ\cos\theta \sqrt{r^2+\ell^2}}{\rho^4}\ .
\end{align}
The homogeneous Maxwell equation is trivially satisfied $F_{[\mu\nu,\sigma]}=0$. 
For the inhomogeneous Maxwell equation we find
\begin{align}
    \nabla^{\hat{\mu}} F_{\hat{\mu} \hat{\nu}} &= J_{\hat{\nu}} = \frac{Q\ell^2}{\rho^7 z} \left(-\frac{\Delta \left(\rho^4+2 \rho^2 z^2-4 z^4\right)}{ z^2 \sqrt{|\Delta|}}, 0, 0, {2 a \sin\theta \left(\rho^2-2 z^2\right)}{}\right).\label{eq:maxinhomo}
\end{align}
We interpret the right-hand side of eq.~\eqref{eq:maxinhomo} as an effective electromagnetic source.
Note that in terms of the (orthonormal) components of the electric and magnetic fields we have $E_{\hat r} = F_{\hat t \hat r}$ and $B_{\hat r} = F_{\hat\theta \hat\phi}$. 
It is then easy to check that this implies that the Maxwell stress-energy tensor \eqref{TMaxwell} is diagonal in this orthonormal basis and that
\begin{align}
    \left[T_\text{Maxwell} \right]^{\hat\mu}{}_{{\hat\nu}} =  \frac{E_{\hat r}^2 + B_{\hat r}^2}{8\pi} \diag\left(-1,-1,1,1\right)\label{TMaxwell_bbKN}
\end{align}
independent of the specific values of $E_{\hat r}$ and $B_{\hat r}$. 
It is also useful to check that
\begin{equation}
    E_{\hat r}^2 +  B_{\hat r}^2
    = \frac{Q^2}{\rho^4} \frac{r^2}{r^2+\ell^2} + \frac{4 Q^2 \ell^2 a^2 \cos^2\theta}{\rho^8}.
\end{equation}

\subsubsection{Interpreting the black-bounce--Kerr--Newman stress-energy}

All of the above treatment is a relatively straightforward generalisation of the Reissner--Nordstr\"om case and also provides the correct limits for $\ell\to 0$ and/or $a\to 0$. 
However, when one attempts to interpret eq.~\eqref{E:xxx} as the sum of the Maxwell stress-energy tensor \eqref{TMaxwell_bbKN} and a charged dust, an inconsistency appears in the form of extra terms. Assuming some generalisation of the energy density of the charged dust and working out the needed electromagnetic potential also does not lead to satisfactory results. 

In what follows, we shall present two alternative interpretations of the stress-energy tensor, one based on a generalisation of the Maxwell dynamics to a non-linear one, the other consisting of a generalisation of the charged dust fluid to one with anisotropic pressure.

\paragraph{Nonlinear electrodynamics}

An alternative to identifying a Maxwell stress-energy tensor in eq.~\eqref{E:xxx} consists in generalising the decomposition of the charged part of the stress energy tensor adopted in the Reissner--Nordstr\"om case to
\begin{equation}\label{TQorth_exp}
    \left[T_Q\right]^{\hat{\mu}}{}_{\hat{\nu}} 
    = \A  \left[T_\text{Maxwell}\right]^{\hat{\mu}}{}_{\hat{\nu}}
    + \Xi\; V^{\hat{\mu}}V_{\hat{\nu}}.
\end{equation}
The multiplicative factor $\A$ will soon be seen to be position-dependent, and to depend on the spin parameter $a$ and regularisation parameter $\ell$, but to be independent of the total charge $Q$. 
This sort of behaviour is strongly reminiscent of nonlinear electrodynamics (NLED) where quite generically one finds $ \left[T_\text{NLED}\right]^{\hat{\mu}}{}_{\hat{\nu}}\propto  \left[T_\text{Maxwell}\right]^{\hat{\mu}}{}_{\hat{\nu}}$.
(For various proposals regarding the use of NLED in regular black hole contexts, see~\cite{AyonBeato:2000,AyonBeato:2004,Lobo:2006,Balart:2014,Rodrigues:2018,Bronnikov:2000,Bolokhov:2012,Bronnikov:2017}.)
The contribution $\Xi\; V^{\hat{\mu}}V_{\hat{\nu}}$ is again that appropriate to charged dust. The 4-velocity $V^{\hat{\mu}}$ is now the (non-geodesic) unit vector parallel to the timelike leg of the tetrad.

If we now compare eq.~\eqref{TMaxwell_bbKN} with $ \left[T_Q\right]^{\hat\mu}{}_{{\hat\nu}}$ as defined in \eqref{TQorth_exp} we identify
\begin{equation}
\A = \frac{Q^2(\rho^2-\ell^2)/\rho^6}{E_{\hat r}^2+B_{\hat r}^2} 
= \frac{\rho^2(\rho^2-\ell^2)(r^2+\ell^2)}{\rho^4 r^2 + 4 a^2\ell^2(r^2+\ell^2) \cos^2\theta}\ .
\end{equation}
We note that at small $\ell$
\begin{equation}
 \A = 1 - \frac{a^2\cos^2\theta \left(3 r^2-a^2\cos^2\theta\right)}{r^2(r^2+a^2\cos^2\theta)^2}\; \ell^2 +\O(\ell^4)    
\end{equation}
So in the limit as $\ell\rightarrow 0$, we see that $\mathcal{A}\rightarrow 1$, restoring standard Maxwell electromagnetism as would be expected for ordinary Kerr--Newman.

Also, we observe the large distance limit
\begin{equation}
\A = 1 -\frac{3\ell^2 a^2 \cos^2\theta}{r^4} + \O(r^{-6}).
\end{equation}
That is, at sufficiently large distances, $ \left[T_Q\right]^{\hat{\mu}}{}_{\hat{\nu}}$ can safely be approximated as a Maxwell-like contribution plus a charged dust, while at small $r$ we have
\begin{equation}
\A = \frac{\ell^2+a^2\cos\theta^2}{4\ell^2}+ \O(r^2).
\end{equation}
This indicates a simple rescaling of the Maxwell stress-energy, (similar to what happens in NLED), deep in the core of the black bounce.

Indeed, it is possible to further characterise the departure from Maxwell-like behaviour by decomposing $\mathcal{A} = 1-a^2\ell^2\mathcal{F}$, where
\begin{equation}
    \mathcal{F} = \frac{\cos^2\theta\left[4(r^2+\ell^2)-{\rho^2}\right]}{\rho^4r^2+ 4a^2\ell^2\cos^2\theta(r^2+\ell^2)}\ .
\end{equation}
The motivation for doing so is to make utterly transparent the correct limiting behaviour for $\mathcal{A}$ both for $a\rightarrow 0$, and for $\ell\rightarrow 0$.

\paragraph{Anisotropic fluid}

As an alternative to the NLED interpretation, we can instead generalise the pressureless dust fluid we had introduced in the Reissner--Nordstr\"om case and impose
\begin{equation}
\left[T_Q \right]_{\hat \mu \hat \nu} -    \left[T_\text{Maxwell} \right]_{\hat \mu \hat \nu} = \diag\left(\varepsilon_{f}, -p_{f}, p_{f}, p_{f}\right),
\label{eq:difference}
\end{equation}
which can be satisfied if 
\begin{equation}
    \varepsilon_{f} = \frac{Q^2 \ell^2}{z^2 \rho^8} \left(4z^4-7z^2\rho^2+\rho^4 \right), \quad p_{f} = \frac{Q^2 \ell^2}{z^2 \rho^8} \left(4z^4-5z^2\rho^2+\rho^4 \right).
\end{equation}
This implies that the right-hand side of eq.~\eqref{eq:difference} can be interpreted, formally, as the stress-energy of an anisotropic fluid.
Specifically, it can be written as
\begin{equation}
    \varepsilon_{f} V_{\hat \mu} V_{\hat \nu} + \frac{p_{f}}{3} \left(g_{\hat \mu \hat \nu} + V_{\hat \mu} V_{\hat \nu} \right) + \pi_{\hat \mu \hat \nu}
\end{equation}
with $V^{\hat \mu} = \left(1,0,0,0\right)$ --- \emph{i.e.}\ $(e_{\hat t})^{\hat \mu}$ --- the velocity of the fluid and \begin{equation}
    \pi_{\hat \mu \hat \nu} = \frac{2p_{f}}{3} \diag\left(0,-2,1,1\right)
\end{equation}
the (traceless) anisotropic shear~\cite{Hawking:1973uf}. 
Note that
\begin{equation}
    p_{f} \propto \left(4z^4-5z^2\rho^2+\rho^4 \right) 
    = \left(4z^2-\rho^2 \right) \left(z^2-\rho^2 \right) \propto a^2 \cos^2\theta.
\end{equation}
So in the limit $a\to 0$ this anisotropic fluid reduces to the usual charged dust.

\section{Discussion and conclusions}\label{S:conclusions}

We have seen that adding an electromagnetic charge to the black-bounce--Schwarz\-schild and black-bounce--Kerr spacetimes leads to the black-bounce--Reissner--Nord\-str\"om and black-bounce--Kerr--Newman spacetimes, which are minimalist, one-pa\-ram\-e\-ter, deformations of the entire Kerr--Newman family which have the desirable properties that they simultaneously (i)~pass all weak-field observational tests, (ii)~are globally regular (no curvature singularities), and (iii)~neatly interpolate between regular black holes and charged traversable wormholes.  

While adding an electromagnetic charge to the black-bounce--Schwarzschild and black-bounce--Kerr spacetimes in this manner is maybe not of direct astrophysical importance (since in any  plausible astrophysical situation  $|Q|/m \ll 1$), it is of considerable theoretical importance, as it gives us an entirely new class of relatively clean everywhere regular black holes to work with. Indeed, we have seen that such geometries present interesting theoretical features such as the existence of a Killing tensor without the presence of the full Killing tower (principal tensor, Killing--Yano tensor) or the fact that the charge-dependent component of the stress-energy for the black-bounce--Kerr--Newman spacetime has a rather non-trivial physical interpretation. In particular, we found that it can either be interpreted as charged dust together with a non-linear modification to standard Maxwell electromagnetism, \emph{or} as standard Maxwell electromagnetism together with an anisotropic fluid. 

While there is no simple way to remove this ambiguity, we can however notice that it appears at least problematic to justify from a physical point of view the introduction of a NLED for the black-bounce--Kerr--Newman spacetime, given that the latter is not required for consistency with general relativity of the black-bounce--Reissner--Nordstr\"om or the black-bounce--Kerr spacetime discussed in ref.~\cite{Liberati:2021}. This seems to suggest that the anisotropic fluid interpretation might be more natural. Overall, we hope that this work will help the understanding of the rich structure of these families of black hole mimickers and stimulate further investigations about their possible realisation in physical settings.

\acknowledgments

EF, SL, and JM acknowledge funding from the Italian Ministry of Education and Scientific Research (MIUR) under the grant PRIN MIUR 2017-MB8AEZ.

\noindent AS acknowledges financial support via a PhD Doctoral Scholarship provided by Victoria University of Wellington. AS is also indirectly supported by the Marsden fund, administered by the Royal Society of New Zealand.

\noindent MV was directly supported by the Marsden Fund, via a grant administered by the Royal Society of New Zealand.

\appendix
\section{Other curvature invariants for the black-bounce--Reissner--Nordstr\"om geometry}\label{S:invariants}

Here we display the explicit forms for the various non-zero curvature invariants for the black-bounce--Reissner--Nordstr\"om geometry. Notably, all non-zero curvature invariants (and orthonormal tensor components) are globally finite, as was immediately guaranteed \emph{via} examination of the Kretschmann scalar in section~\ref{SS:Kretschmann}.

For the Ricci scalar
\begin{equation}
    R = -\frac{2\ell^2}{(r^2+\ell^2)^{3}}\left[Q^2+r^2+\ell^2-3m\sqrt{r^2+\ell^2}\right] \ ,
\end{equation}
and as $r\rightarrow 0$
\begin{equation}
    \lim_{r\rightarrow 0} R = -\frac{2(Q^2+\ell^2-3m\ell)}{\ell^4} \ .
\end{equation}
The quadratic Ricci invariant $R^{\mu\nu}R_{\mu\nu}$ is
\begin{align}
    R^{\mu\nu}R_{\mu\nu} &= \frac{4\left(r^4-\ell^2r^2+\ell^4\right)}{(r^2+\ell^2)^{6}}Q^{4} + \frac{4\ell^2\left[m(r^2-4\ell^2)+(r^2+\ell^2)^{3/2}\right]}{(r^2+\ell^2)^{11/2}}Q^2 \nonumber \\
    & + \frac{2\ell^4\left[2r^4-6m(r^2+\ell^2)^{3/2}+(9m^2+4\ell^2)(r^2+\ell^2)-2\ell^4\right]}{(r^2+\ell^2)^{6}} \ ,
\end{align}
and as $r\rightarrow 0$ we see
\begin{equation}
    \lim_{r\rightarrow 0} R^{\mu\nu}R_{\mu\nu} = \frac{4}{\ell^8}\,Q^4+\frac{4(\ell-4m)}{\ell^7}\,Q^2+\frac{4\ell^2-12m\ell+18m^2}{\ell^6} \ .
\end{equation}
For the Weyl contraction $C^{\mu\nu\rho\sigma}C_{\mu\nu\rho\sigma}$ we obtain
\begin{align}
    &
    C^{\mu\nu\rho\sigma}C_{\mu\nu\rho\sigma} \nonumber\\
    &\quad= \frac{16(3r^2-\ell^2)^{2}}{3(r^2+\ell^2)^{6}}Q^4 + \frac{16(\ell^2-3r^2)\left[2\ell^2\sqrt{r^2+\ell^2}-3m(\ell^2-2r^2)\right]}{3(r^2+\ell^2)^{11/2}}Q^2 + \frac{48m^2r^6}{(r^2+\ell^2)^{6}} \nonumber \\
    &\quad + \frac{4\ell^4\left[4\ell^4+(9m^2+8r^2)\ell^2+4r^4-27m^2r^2-12m\sqrt{r^2+\ell^2}(1+\frac{r^2}{\ell^2})(\ell^2-2r^2)\right]}{3(r^2+\ell^2)^{6}} \ ,
\end{align}
and as $r\rightarrow 0$
\begin{equation}
    \lim_{r\rightarrow 0}C^{\mu\nu\rho\sigma}C_{\mu\nu\rho\sigma} = \frac{4\left(2Q^2+2\ell^2-3\ell m\right)^2}{3\ell^8} \ .
\end{equation}

\bibliographystyle{JHEP.bst}
\bibliography{references}

\end{document}